\documentclass[prl,twocolumn]{revtex4}
\usepackage{graphicx, epsfig}
\usepackage{color}
\usepackage{mathrsfs}

\newcommand{\be}{\begin{equation}}
\newcommand{\ee}{\end{equation}}
\newcommand{\bea}{\begin{eqnarray}}
\newcommand{\eea}{\end{eqnarray}}
\newcommand{\ba}{\begin{eqnarray}}
\newcommand{\ea}{\end{eqnarray}}

\newcommand{\gapp}{\mathrel{\raise.3ex\hbox{$>$}\mkern-14mu
              \lower0.6ex\hbox{$\sim$}}}
\newcommand{\lapp}{\mathrel{\raise.3ex\hbox{$<$}\mkern-14mu
              \lower0.6ex\hbox{$\sim$}}}

\begin{document}
\title{From Particles to Kinks}

\author{Sourish Dutta}
\affiliation{
Department of Physics and Astronomy, Vanderbilt University, 
Nashville, TN 37235}
\author{D.~A.~Steer}
\affiliation{
APC, UMR 7164, 10 rue Alice Domon et L\'eonie Duquet,
75205 Paris Cedex 13, France
}
\author{Tanmay Vachaspati}
\affiliation{
Institute for Advanced Study, Princeton, NJ 08540\\ 
CERCA, Department of Physics, 
Case Western Reserve University, Cleveland, OH~~44106-7079
}

%%%%%%%%%%%%%%%%%%%%%%%%%%%%%%%%%%%%%%%%%%%%%%%%%%%%%%%
\begin{abstract}
\noindent
We study the creation of solitons from particles, using the
$\lambda \phi^4$ model as a prototype. We consider the scattering 
of small, identical, wave pulses, that are equivalent to a 
sequence of particles, and find that kink-antikink pairs
are created for a large region in parameter space. We also
find that scattering at {\it low} velocities is favorable 
for creating solitons that have large energy compared to the 
mass of a particle. 
\end{abstract}
%%%%%%%%%%%%%%%%%%%%%%%%%%%%%%%%%%%%%%%%%%%%%%%%%%
%\pacs{???}

\maketitle

A wide variety of systems, ranging from polyacetylene and 
Josephson junctions to high energy particle physics models,
contain non-perturbative, ``soliton'' or ``solitary
wave'' excitations in addition to pertubative ``particle'' 
excitations \cite{Vachaspatibook}. An important unsolved 
problem is to find ways to transition from the particle 
sector to the soliton sector. At a pragmatic level, we would 
like to develop implementable schemes that might enable solitons 
to be built out of particles. A transition from two energetic 
particles to solitons, however, is known to be exponentially 
suppressed e.g.\ \cite{Levkov:2004tf} (for a review see
\cite{Mattis:1991bj}) though  it
may occur more readily in certain situations, 
such as in the background of a pre-existing kink,\cite{Manton:1996ex,Romanczukiewicz:2005rm}.

In this paper, we will determine a class of initial conditions 
that consist of small amplitude perturbations that scatter and
successfully lead to the production of a kink-antikink 
(``${\rm k {\bar k}}$'') pair in 1+1 dimensions. A trivial 
scheme to determine such a set of initial conditions is to 
time reverse the annihilation of ${\rm k {\bar k}}$. Then 
the time reversed particles 
would assemble into an outgoing ${\rm k {\bar k}}$. 
However, in any practical setting, such initial conditions 
would be very hard to arrange since the characteristics
of the radiation from ${\rm k {\bar k}}$ annihilation 
are highly non-trivial. Instead we want to consider 
``clean'' initial conditions in which we scatter identical 
wave pulses, somewhat like 2 particle scattering.
The simplicity of the initial state comes with a price 
in that the final state will now not only contain a 
${\rm k {\bar k}}$ but also some radiation. Our approach 
thus differs from other studies which generally considered 
initial conditions containing a single kink and hence had 
non-vanishing topological charge, e.g.~\cite{Romanczukiewicz:2005rm}.

In order to determine clean initial conditions that give
${\rm k {\bar k}}$ in the final state, we draw lessons from the
sine-Gordon model which contains both particle and soliton
sectors and has been studied extensively, 
both classically and in quantum theory 
\cite{Rajaramanbook,Colemanbook,Vachaspatibook,
Dashen:1975hd}.
While the complete integrability 
of the sine-Gordon model permits many exact solutions,
it also leads to a disappointing disconnection between the 
particle and soliton sectors, not present in many other 
models which admit solitons. For example, in the sine-Gordon 
model, it is not possible to start with, say, a soliton and 
an antisoliton and end up with particles. If a soliton and 
an antisoliton are set up to collide and possibly annihilate, 
they simply pass through each other. Thus soliton scattering 
states do not convert to particle states (even in quantum
theory).

What is important for us is that the sine-Gordon model also 
contains ``breather states''. If a breather state has large
amplitude, it can be interpreted as a bound state of a soliton 
and an antisoliton, in which the two keep oscillating about each 
other but never annihilate.  
On the other hand, small quantized breathers have been 
interpreted as fundamental particles in the theory. Then 
one might expect the breather to be a bridge between the 
particle and soliton sectors. In the sine-Gordon model, 
however, the breather is a stable object in itself and fails 
to connect the particle and soliton sectors.

To connect the particle and soliton sectors it is necessary 
to depart from the sine-Gordon model. The smaller the departure,
the weaker will be the connection between the particle and soliton
sectors. Then, if we depart weakly from the sine-Gordon model,
we expect long-lived ``breather-like'' states that can transition
to both widely separated kink-antikink pair and also to particles. 
Such long-lived states have been discovered in various systems
and have been termed ``bions'' in certain contexts and ``oscillons'' 
in others \cite{Bogolyubsky:1976nx,boyd1,Gleiser:1993pt,Copeland:1995fq,
boyd2,Honda:2000gv,Farhi:2005rz,Hindmarsh:2006ur,Fodor:2006zs,
Saffin:2006yk,Graham:2006vy,Gleiser:2007te,Fodor08}.

Motivated by these considerations, we study ${\rm k {\bar k}}$ 
production in the $\lambda \phi^4$ model
\begin{equation}
L = \phi_0^2 \left [ \frac{1}{2} (\partial_\mu \phi )^2 - 
                  \frac{1}{4} \left ( \phi^2 - 1 \right )^2
              \right ]
\label{modelL}
\end{equation}
where we have rescaled fields and coordinates so that $\phi_0$
is the only parameter in the model. 
%Further, $\phi_0$ does
%not affect the classical dynamics, though it does play a
%role in the quantized model. 
%
The equation of motion is
\begin{equation}
{\ddot \phi} = \phi'' - (\phi^2 -1) \phi
\label{eom}
\end{equation}
where overdots denote time derivatives and primes denote
spatial derivatives.
The mass of a fundamental excitation can be found by considering
small fluctuations around one of the vacua (say $\phi=+1$)
and is $m = \sqrt{2}$. The kink profile is
\begin{equation}
\phi_k = {\rm tanh}\left ( \frac{x}{\sqrt{2}} \right )
\end{equation}
The energy of a kink is found from the energy expression
\begin{equation}
E = \phi_0^2 \int dx \left [ \frac{\dot\phi ^2}{2}
                    + \frac{{\phi '} ^2}{2}
                    + \frac{1}{4}(\phi^2-1)^2 \right ]
\label{energyexpression}
\end{equation}
and is 
\begin{equation}
E_{\rm k} = \frac{2\sqrt{2}}{3} \phi_0^2 
          = \frac{2m}{3} \phi_0^2 
\end{equation}
Note that the kink energy may be made very large compared 
to the particle mass by taking large values of $\phi_0$.
However, $\phi_0$ itself does not enter the classical dynamics of
the scalar field  though it does play a r\^ole in the quantized model. 

We would like to use breather-like solutions in the
$\lambda\phi^4$ model in our initial condition. 
However, such solutions are not known analytically.  Hence we
simply use the breather solutions of the pure sine-Gordon model 
\begin{equation}
 L _{sG} =  \frac{1}{2} \partial_\mu \phi \partial^\mu \phi - 
           \frac{1}{\pi^2} [ 1+\cos (\pi \phi)]
 \end{equation}
 which are given by
\begin{equation}
\phi^{sG}_b (t,x; \omega , v) = -1+ 
    \frac{4}{\pi} \tan^{-1} \left [ \frac{\eta \sin(\omega T)}
                            {\cosh (\eta \omega X)} \right ]
\label{breather}
\end{equation}
where
\begin{eqnarray}
T &=& \gamma [t - v (x-x_0)] \ , \ \ X = \gamma [x-x_0 - v t]
\nonumber \\
\gamma &=& (1-v^2)^{-1/2} \ ,  \ \ \eta = \sqrt{1-\omega^2} /\omega  \, .
\end{eqnarray}

%If we set up sine-Gordon breather initial conditions 
%but in the $\lambda \phi^4$ model, we can check 
%that this gives a quasi-breather solution that oscillates 
%for a long time before decaying.
%A boosted, shifted, breather solution in the sine-Gordon
%model is
%\begin{equation}
%\phi^{sG}_b (t,x; \omega , v) = -1+ 
%    \frac{4}{\pi} \tan^{-1} \left [ \frac{\eta \sin(\omega T)}
%                            {\cosh (\eta \omega X)} \right ]
%\label{breather}
%\end{equation}
%where
%\begin{eqnarray}
%T &=& \gamma [t - v (x-x_0)] \ , \ \ X = \gamma [x-x_0 - v t]
%\nonumber \\
%\gamma &=& (1-v^2)^{-1/2} \ ,  \ \ \eta = \sqrt{1-\omega^2} /\omega  
%\end{eqnarray}
In Eq.~(\ref{breather}), the $\tan^{-1}(\cdot)$ function is taken to lie in the interval
$(-\pi /2,+\pi/2)$.
Apart from the boost $\gamma$ and shift $x_0$, a breather solution is
labelled by the parameter $\omega \in (0,1)$. The 
solution for small $\omega$ can be viewed as a sine-Gordon 
kink and an antikink that are oscillating back and forth, merging 
and emerging forever. Note that the breather is localized around 
one vacuum (at $-1$), and probes the second vacuum (at $+1$) 
for durations that vary inversely with $\omega$.
For $\omega \approx 1$, the breather describes oscillations in 
the vacuum around $\phi = -1$. In the quantum theory, these
oscillations are quantized and the energy of the lowest quantum 
state is equal to that of a particle, leading to the identification
of the lowest energy breather with the particle excitation in
the model.

Before proceeding 
%it is important to understand the evolution of 
consider an initial unboosted sine-Gordon breather (Eq.~(\ref{breather}) 
with $v=0$), in the $\lambda \phi^4$ model with equation of motion given in Eq.~(\ref{eom}).  (That is, the initial condition is $\phi(0,x)=\phi^{sG}_b (0,x; \omega , 0)$ and $\dot{\phi}(0,x)=\dot{\phi}^{sG}_b (0,x; \omega , 0)$.)
The energy of the solution can be obtained by evaluating Eq.~(\ref{energyexpression}) at $t=0$ when $\phi_b =-1$ for 
all $x$. Then the potential and gradient terms do not contribute, and the kinetic 
contribution is easily evaluated. As in the sine-Gordon
model we find
\begin{equation}
E_{\rm b} = \frac{16}{\pi^2} \sqrt{1-\omega^2} \phi_0^2
\end{equation}
The ratio of kink to breather energy is
\begin{equation}
\frac{E_{\rm k}}{E_{\rm b}} = \frac{\pi^2}{12\sqrt{2}}
      \frac{1}{\sqrt{1-\omega^2}} \approx 
          \frac{\pi^2}{24} \frac{1}{\sqrt{1-\omega}}
\label{ktobratio}
\end{equation}
where in the last expression we assume $\omega \approx 1$.  
The field profile itself, $\phi(t,x)$, can be obtained numerically 
and we have checked that it is oscillatory and long-lived.  
More specifically, we have shown that half of the initial 
energy $E_{\rm b}$ in the simulation box (itself much larger 
than the breather size) is radiated in a time 
$T_{1/2} \simeq 5 \times 10^{4} \lambda^{-1.9}$, 
independently of $\omega$.

We now turn to the problem at hand, namely the creation of 
${\rm k {\bar k}}$ from particles.  Our initial conditions will 
consist of a train of $N_b$ 
little ({\it i.e.} $\omega \approx 1$) breathers coming 
in from the left and another identical train of $N_b$ 
breathers coming in from the right. We will study the
collision of these breather trains for a variety of
parameters and look for the formation of ${\rm k {\bar k}}$. Hence
our initial condition corresponds to an incoming state
\begin{equation}
f(t,x) = -1 + 
   \hskip -0.1 cm \sum_{n=-N_b, \ne 0}^{N_b} \frac{4}{\pi} 
 \tan^{-1} \left [ \frac{\eta \sin(\omega T_n)}
                            {\cosh (\eta \omega X_n)} \right ]
\label{initialcondition}
\end{equation}
with
\begin{equation}
T_n = \gamma [t - v_n (x-x_{0n} )] \ , \ \ 
X_n = \gamma [(x-x_{0n}) - v_n t]
\end{equation}
where $x_{0n} = a+nd$, $ v_n = - v <0$ for $n >0$, and
$x_{0n}=-a+nd$, $v_n = +v >0$ for $n<0$. The parameter $a$ 
is half the separation between the trains at $t=0$ and $d$ 
is the separation between different breathers in the same 
train. The initial conditions (at $t=0$) are 
\begin{equation}
\phi (0,x) = f(0,x) \ , \ \ {\dot \phi} (0,x) = {\dot f}(0,x)
\end{equation}

To further motivate our choice of initial conditions,
let us consider what might be required to form a ${\rm k {\bar k}}$.
Initially, the field is oscillating about the $\phi =-1$
vacuum. To form ${\rm k {\bar k}}$, we need the oscillations to 
extend into the $\phi=+1$ vacuum. So we need to build up the field
oscillations. As the leading breathers in the
trains collide, the field at $x=0$ starts oscillating.
Subsequent breathers provide additional kicks to the 
oscillations at $x=0$. Provided the subsequent collisions 
are in phase, the amplitude of oscillations at $x=0$ will
grow in resonance. The growth must compete with the 
dissipation due to the emission of particle radiation.
If the growth wins, the oscillations will eventually extend up 
to the $\phi=+1$ vacuum and then it will be energetically 
favorable for $\phi(t,0)$ to stay there. Then it becomes 
likely that ${\rm k {\bar k}}$ will be created.

The same heuristic argument may be applied to the pure 
sine-Gordon model and serves to show its limitations. We know 
that ${\rm k {\bar k}}$ are not created in the sine-Gordon model, 
but it is not because dissipation is stronger than resonant
growth. Instead the integrability ensures that the
breather trains pass unscathed through each other.
So the heuristic argument should be taken as motivation
but cannot be taken too literally; instead we must solve
the equations of motion and check for ${\rm k {\bar k}}$ production.
However, what seems clear is that ${\rm k {\bar k}}$ production may
proceed via a resonance and, just as a child can swing
higher and higher by timing her movement to within a
factor of 2 per kick, this level of tuning may be all 
that is needed to produce ${\rm k {\bar k}}$.

The equation of motion for the scalar field, Eq.~(\ref{eom}), 
is solved numerically with the initial conditions in 
Eq.~(\ref{initialcondition}), using the iterated Crank-Nicholson 
method with two iterations \cite{Teukolsky:1999rm}, and
absorbing boundary conditions \cite{Olum:1999sg}
at the ends of the lattice. The fields are evolved for one 
light crossing time. As an additional check, we have also 
evolved the intial conditions using Mathematica, though with 
fixed boundary conditions (see \cite{webmovie} for the notebook). 
The Mathematica results are generally consistent with the 
Crank-Nicholson method but there are a few discrepancies. These 
may be due to the different integration routines which also have 
different accuracies. The Crank-Nicholson implementation is more 
transparent and we find it more reliable, while the Mathematica
implementation is more convenient to use.

The problem contains many parameters, all related to
the choice of initial condition: $\omega$, $v$, $a$, 
$d$, and $N_b$. For a given value of $\omega$, 
Eq.~(\ref{ktobratio}) shows that, just on energetic grounds, 
we need $N_b > 0.6/\sqrt{1-\omega^2}$. We have taken
$N_b = {\rm int}[2/\sqrt{1-\omega^2}]+1$ where 
${\rm int}[x]$ denotes the largest integer less than or 
equal to $x$. Somewhat arbitrarily,
we take the initial half-separation of the trains to be 
$a = 10/\eta \omega =10/\sqrt{1-\omega^2}$, corresponding
to 10 (at rest) breather widths. 
The separation of the breathers in a train is taken to be
$d=2/\sqrt{1-\omega^2}$. The only parameters
left to specify are $\omega$ and $v$.
For a breather to have energy comparable to the particle
mass, and the kink to have energy much larger than the
particle, we require $\omega$ to be very close to 1.
With $\omega =0.99$, the kink energy is about 4 times
that of a breather. We shall take $\omega \in (0.90,0.99)$. 
We then do runs for different values of $v$ and look 
for ${\rm k {\bar k}}$ formation.

\begin{figure}
  \includegraphics[width=3.2in,angle=0]{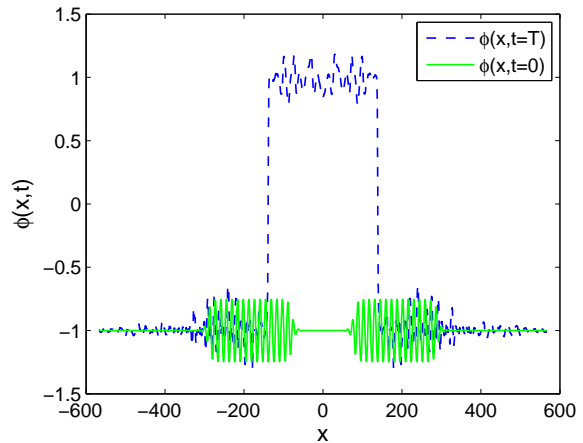}
\caption{Two snapshots of the collision of breather trains for 
$\omega=0.99$, $v=0.43$ and other parameters as described in the 
text.  $T$ denotes a light crossing time.  The initial
state contains the train of breathers. Subsequently,
kinks appear and move apart.
}
\label{trainevolution}
\end{figure}

An example of ${\rm k {\bar k}}$ production is shown in 
Fig.~\ref{trainevolution} where we give two snapshots of 
the evolution. Animations of the evolution may be found in 
Ref.~\cite{webmovie}. 
Generally, by looking at the field profile, it is quite 
clear when a ${\rm k {\bar k}}$ has been created. However,
there are some instances in which the outcome is not so
clear-cut. This includes the case when the field profile
shows ${\rm k {\bar k}}$ that are not separated by a large
distance or are almost at rest with respect to each other. 
Then there is the possibility that the ${\rm k {\bar k}}$
will annihilate. In such cases, we have chosen to call it
a ${\rm k {\bar k}}$ creation event if the kinks survived 
for at least the duration of the simulation. Another
novel outcome we have seen is that for some parameters 
two or more pairs of ${\rm k {\bar k}}$ are produced.

\begin{figure}
\hskip -0.25 cm  \includegraphics[width=3.0in,angle=-90]{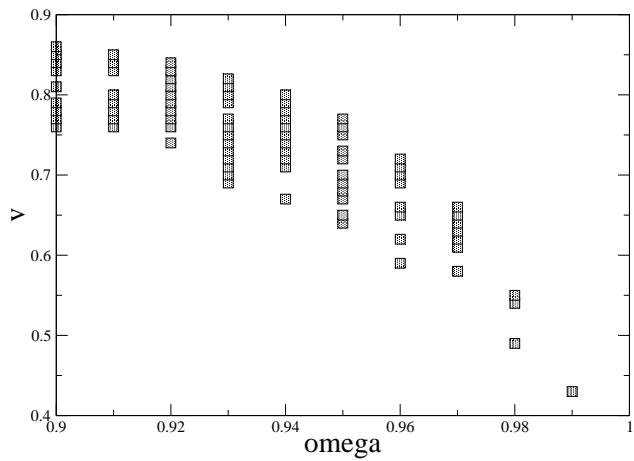}
\caption{  
Results from our Crank-Nicholson code (square symbols) mapping
out  ${\rm k {\bar k}}$ formation in the $(\omega , v)$ plane 
for the choice of other parameters as described in the text.
Note the occasional gaps where ${\rm k {\bar k}}$ are not
formed, and the downward trend with larger $\omega$ (weaker
incoming pulses).  
}
\label{parameterregion}
\end{figure}

In Fig.~\ref{parameterregion} we plot the region on
the $(\omega, v)$ plane for our choice of parameters
that lead to the formation of a pair (or more) of
${\rm k {\bar k}}$. Note the trend -- higher $\omega$
{\it i.e.} smaller breather energy, requires lower
incoming velocity. This indicates that it is preferable
to scatter many particles at {\em low} energy to create 
solitons. If the incoming velocity is too
high, the breather trains simply pass through, as 
in the sine-Gordon model. Also, note the occasional
holes in the plot ({\it e.g.} $\omega=0.91$, $v=0.82$) 
where we did not observe ${\rm k {\bar k}}$ formation. 
This substructure in the plot is reminiscent of the
bands observed in ${\rm k {\bar k}}$ {\it scattering} 
\cite{Matzner} and suggests that ${\rm k {\bar k}}$ 
formation may be due to resonance.

The region leading to 
${\rm k {\bar k}}$ formation is reasonably large
but does not extend to arbitrarily high $\omega$.
For example, we have not found initial conditions
leading to ${\rm k {\bar k}}$ formation for
$\omega > 0.99$. The expanse of the ``successful'' region 
does not concern us at the moment because our main 
objective was to find a set of 
clean initial conditions that led to the formation of 
${\rm k {\bar k}}$. We would be surprised if future 
investigations do not find a larger set of successful
clean initial conditions, even for very high values of 
$\omega$. Whether these initial conditions are achievable 
in a practical setting is a separate matter, and depends
on the details of the experiment. 

There are several directions in which it would be useful 
to extend our results. The first is to scan the space
of initial conditions more carefully, to gain further
understanding of what conditions enable ${\rm k {\bar k}}$
formation. Our space of initial conditions could also be 
enlarged, if necessary. For example, different breathers 
in a train could come in with different velocities. We could 
also envision ``building up'' by starting with very large 
$\omega$ (small energy) breathers, and building
states corresponding to smaller $\omega$ (larger energy), 
which can then collide to form ${\rm k {\bar k}}$.
Another direction is to include quantum effects in 
the scattering. This would require more precise
understanding of the breather and kink states in terms of
particles. In the quantum sine-Gordon model, soliton operators 
have been written down in terms of an infinite number of 
particle operators \cite{Mandelstam:1975hb}. We expect that 
the soliton operator in the $\lambda \phi^4$ model should 
be expressible in terms of a finite number of particle
operators otherwise it would seem impossible to build
a ${\rm k {\bar k}}$ starting with particles.  Yet another 
direction to proceed would be to consider solitons in higher 
dimensions. Then we can study the creation of vortex-antivortex 
or monopole-antimonopole pairs in
suitable systems. We would clearly need higher dimensional
analogs of breathers and we expect that oscillon states 
can play this role. Finally, it would be useful to
generalize our initial state to real systems. After all,
polyacetylene is described by the $\lambda \phi^4$  
model and we may expect to be able to create 
${\rm k {\bar k}}$ there. (Similar problems also arise 
in polymer physics in the context of polymers that
pass through a membrane \cite{SebPau2000}.) Our results 
do not directly apply to polyacetylene because the dynamics 
there is non-relativistic. However, with suitable 
generalization, it may become possible to test some of 
these ideas experimentally. 
 
SD and DAS thank the Institute for Advanced Study for
hospitality and computer support. DAS is grateful to 
Nino Flachi for discussion. TV thanks Nima Arkani-Hamed,
Peter Goldreich, Simeon Hellerman, Juan Maldacena, 
Nathaniel Seiberg, Tomer Volansky, and Edward Witten 
for suggestions and comments.
This work was supported by the U.S. Department of Energy 
and NASA at Case Western Reserve University.


\begin{thebibliography}{9}


\bibitem{Vachaspatibook}
``Kinks and Domain Walls'',
T. Vachaspati, Cambridge University Press (2006).

%\cite{Levkov:2004tf}
\bibitem{Levkov:2004tf}
  D.~G.~Levkov and S.~M.~Sibiryakov,
  %``Induced tunneling in QFT: Soliton creation in collisions of highly
  %energetic particles,''
  Phys.\ Rev.\  D {\bf 71}, 025001 (2005)
  [arXiv:hep-th/0410198].
  %%CITATION = PHRVA,D71,025001;%%

%\cite{Mattis:1991bj}
\bibitem{Mattis:1991bj}
  M.~P.~Mattis,
  %``The Riddle of high-energy baryon number violation,''
  Phys.\ Rept.\  {\bf 214}, 159 (1992).
  %%CITATION = PRPLC,214,159;%%

%\cite{Manton:1996ex}
\bibitem{Manton:1996ex}
  N.~S.~Manton and H.~Merabet,
  %``phi~4 Kinks - Gradient Flow and Dynamics,''
  arXiv:hep-th/9605038.
  %%CITATION = HEP-TH/9605038;%%

%\cite{Romanczukiewicz:2005rm}
\bibitem{Romanczukiewicz:2005rm}
  T.~Romanczukiewicz,
  %``Creation of kink and antikink pairs forced by radiation,''
  J.\ Phys.\ A  {\bf 39}, 3479 (2006)
  [arXiv:hep-th/0501066].
  %%CITATION = JPAGB,A39,3479;%%

\bibitem{Rajaramanbook}
``Solitons and Instantons'',
R. Rajaraman, North-Holland, Amsterdam (1987).

\bibitem{Colemanbook}
``Aspects of Symmetry: Selected Erice Lectures'',
S. Coleman, Cambridge University Press (1985).

%\cite{Dashen:1975hd}
\bibitem{Dashen:1975hd}
  R.~F.~Dashen, B.~Hasslacher and A.~Neveu,
  %``The Particle Spectrum In Model Field Theories From Semiclassical Functional
  %Integral Techniques,''
  Phys.\ Rev.\  D {\bf 11}, 3424 (1975);
  %%CITATION = PHRVA,D11,3424;%%
%\cite{Dashen:1974cj}
%\bibitem{Dashen:1974cj}
  %R.~F.~Dashen, B.~Hasslacher and A.~Neveu,
  %%``Nonperturbative Methods And Extended Hadron Models In Field Theory. 2.
  %%Two-Dimensional Models And Extended Hadrons,''
  Phys.\ Rev.\  D {\bf 10}, 4130 (1974);
  %%%CITATION = PHRVA,D10,4130;%%
%%\cite{Dashen:1974ci}
%\bibitem{Dashen:1974ci}
  %R.~F.~Dashen, B.~Hasslacher and A.~Neveu,
  %%``Nonperturbative Methods And Extended Hadron Models In Field Theory. 1.
  %%Semiclassical Functional Methods,''
  Phys.\ Rev.\  D {\bf 10}, 4114 (1974).
  %%%CITATION = PHRVA,D10,4114;%%

%%\cite{Mattis:1991bj}
%\bibitem{Mattis:1991bj}
  %M.~P.~Mattis,
  %%``The Riddle of high-energy baryon number violation,''
  %Phys.\ Rept.\  {\bf 214}, 159 (1992).
  %%%CITATION = PRPLC,214,159;%%

%\cite{Bogolyubsky:1976nx}
\bibitem{Bogolyubsky:1976nx}
  I.~L.~Bogolyubsky and V.~G.~Makhankov,
  %``On The Pulsed Soliton Lifetime In Two Classical Relativistic Theory
  %Models,''
  JETP Lett.\  {\bf 24}, 12 (1976).
  %%CITATION = JTPLA,24,12;%%

\bibitem{boyd1}  J.~P.~Boyd, Nonlinearity {\bf 3} (1990) 177-195.

%\cite{Gleiser:1993pt}
\bibitem{Gleiser:1993pt}
  M.~Gleiser,
  %``Pseudostable bubbles,''
  Phys.\ Rev.\  D {\bf 49}, 2978 (1994)
  [arXiv:hep-ph/9308279].
  %%CITATION = PHRVA,D49,2978;%%

%\cite{Copeland:1995fq}
\bibitem{Copeland:1995fq}
  E.~J.~Copeland, M.~Gleiser and H.~R.~Muller,
  %``Oscillons: Resonant configurations during bubble collapse,''
  Phys.\ Rev.\  D {\bf 52}, 1920 (1995)
  [arXiv:hep-ph/9503217].
  %%CITATION = PHRVA,D52,1920;%%

\bibitem{boyd2} J.~P.~Boyd, Wave Motion {\bf 21} (1995) 311-330.

%\cite{Honda:2000gv}
\bibitem{Honda:2000gv}
  E.~P.~Honda,
  %``Resonant dynamics within the nonlinear Klein-Gordon equation: Much ado
  %about oscillons,''
  arXiv:hep-ph/0009104.
  %%CITATION = HEP-PH/0009104;%%

%\cite{Farhi:2005rz}
\bibitem{Farhi:2005rz}
  E.~Farhi, N.~Graham, V.~Khemani, R.~Markov and R.~Rosales,
  %``An oscillon in the SU(2) gauged Higgs model,''
  Phys.\ Rev.\  D {\bf 72}, 101701 (2005)
  [arXiv:hep-th/0505273].
  %%CITATION = PHRVA,D72,101701;%%

%\cite{Hindmarsh:2006ur}
\bibitem{Hindmarsh:2006ur}
  M.~Hindmarsh and P.~Salmi,
  %``Numerical investigations of oscillons in 2 dimensions,''
  Phys.\ Rev.\  D {\bf 74}, 105005 (2006)
  [arXiv:hep-th/0606016].
  %%CITATION = PHRVA,D74,105005;%%

%\cite{Fodor:2006zs}
\bibitem{Fodor:2006zs}
  G.~Fodor, P.~Forgacs, P.~Grandclement and I.~Racz,
  %``Oscillons and quasi-breathers in the phi**4 Klein-Gordon model,''
  Phys.\ Rev.\  D {\bf 74}, 124003 (2006)
  [arXiv:hep-th/0609023].
  %%CITATION = PHRVA,D74,124003;%%

%\cite{Saffin:2006yk}
\bibitem{Saffin:2006yk}
  P.~M.~Saffin and A.~Tranberg,
  %``Oscillons and quasi-breathers in D+1 dimensions,''
  JHEP {\bf 0701}, 030 (2007)
  [arXiv:hep-th/0610191].
  %%CITATION = JHEPA,0701,030;%%

%\cite{Graham:2006vy}
\bibitem{Graham:2006vy}
  N.~Graham,
  %``An Electroweak Oscillon,''
  Phys.\ Rev.\ Lett.\  {\bf 98}, 101801 (2007)
  [arXiv:hep-th/0610267].
  %%CITATION = PRLTA,98,101801;%%

%\cite{Gleiser:2007te}
\bibitem{Gleiser:2007te}
  M.~Gleiser and J.~Thorarinson,
  %``A phase transition in U(1) configuration space: Oscillons as remnants of
  %vortex antivortex annihilation,''
  arXiv:hep-th/0701294.
  %%CITATION = HEP-TH/0701294;%%

\bibitem{Fodor08}
 G.~Fodor, P.~Forgacs, Z.~Horvath and A.~Lukacs,
 arXiv:0802.3525.
 
\bibitem{Matzner}
  P.~Anninos, S.~Oliveira and R.~A.~Matzner,
  %``Fractal structure in the scalar lambda (phi**2-1)**2 theory,''
  Phys.\ Rev.\  D {\bf 44} (1991) 1147.
  %%CITATION = PHRVA,D44,1147;%% 
 

%\cite{Teukolsky:1999rm}
\bibitem{Teukolsky:1999rm}
  S.~A.~Teukolsky,
  %``On the Stability of the Iterated Crank-Nicholson Method in Numerical
  %Relativity,''
  Phys.\ Rev.\  D {\bf 61}, 087501 (2000)
  [arXiv:gr-qc/9909026].
  %%CITATION = PHRVA,D61,087501;%%

\bibitem{webmovie}
http://www.hep.vanderbilt.edu/~duttas/kkbar.html

%\cite{Olum:1999sg}
\bibitem{Olum:1999sg}
  K.~D.~Olum and J.~J.~Blanco-Pillado,
  %``Radiation from cosmic string standing waves,''
  Phys.\ Rev.\ Lett.\  {\bf 84}, 4288 (2000)
  [arXiv:astro-ph/9910354].
  %%CITATION = PRLTA,84,4288;%%

%\cite{Mandelstam:1975hb}
\bibitem{Mandelstam:1975hb}
  S.~Mandelstam,
  %``Soliton operators for the quantized sine-Gordon equation,''
  Phys.\ Rev.\  D {\bf 11}, 3026 (1975).
  %%CITATION = PHRVA,D11,3026;%%

\bibitem{SebPau2000}
K.L.~Sebastian and A.K.R.~Paul,
%Kramers problem for a polymer in a double well
Phys.\ Rev.\ E, {\bf 62}, 927 (2000).



\end{thebibliography}
\end{document}